\documentclass[10pt,conference]{IEEEtran}

\usepackage[a4paper,margin=1in]{geometry}
\usepackage{times}
\usepackage[english]{babel}
\usepackage{hyperref}
\usepackage{amsmath,amssymb}
\usepackage{comment}
\usepackage{blindtext}
\usepackage{mathtools}
\usepackage{adjustbox}
\usepackage{multirow}
\usepackage{makecell}
\usepackage[inline]{enumitem}
\usepackage{longtable}
\usepackage{tabularray}
\usepackage{tabularx}
\usepackage[dvipsnames]{xcolor}
\usepackage{tikz}
\usepackage{pgfplots}
\pgfplotsset{compat=1.18}

\newif\ifstatus
\statustrue

\usepackage{lineno}

\newcommand{\pockets}{{POK}}
\newcommand{\commontopic}{\textit{common topic}}
\newcommand{\commontopics}{\textit{common topics}}
\newcommand{\uniquetopic}{\textit{unique topic}}
\newcommand{\uniquetopics}{\textit{unique topics}}
\newcommand{\othertext}{\textit{non-\-con\-ver\-sa\-tio\-nal text}}

\title{Investigating and Comparing Discussion Topics in Multilingual Underground Forums}

\author{
\IEEEauthorblockN{Mariella Mischinger}
\IEEEauthorblockA{
IMDEA Networks Institute, Madrid, Spain\\
Universidad Carlos III de Madrid, Leganés, Spain
}
\and
\IEEEauthorblockN{Vahid Ghafouri}
\IEEEauthorblockA{
IMDEA Networks Institute, Madrid, Spain\\
Universidad Carlos III de Madrid, Leganés, Spain
}
\and
\IEEEauthorblockN{Sergio Pastrana}
\IEEEauthorblockA{
Universidad Carlos III de Madrid, Leganés, Spain
}
\and
\IEEEauthorblockN{Guillermo Suarez-Tangil}
\IEEEauthorblockA{
IMDEA Networks Institute, Madrid, Spain
}
}

\begin{document}

\maketitle

\begin{abstract}
Underground forums play a crucial role in the criminal ecosystem, facilitating the exchange of knowledge and the trade of illegal tools and services.
By analyzing the skills, motivations, focus, and operations of cyber-criminals active in these forums, cybersecurity professionals and law enforcement can better understand their tactics, assess the risks they pose to society, and develop more effective countermeasures. 
A significant challenge in analyzing these forums arises from language barriers, either because they blend different languages or because they use community-specific slang. 

In this paper, we address this challenge through the use of a  combination of unsupervised methods that group together semantically related conversational themes (i.e., topics) into clusters. 
We apply our methodology to analyze a prolific, invite-only, Russian-English criminal forum that has been operating for over 18 years. 
This way, we uncover \textit{pockets of knowledge}, i.e., knowledge only shared in one sub-community. 
This knowledge is accessible only to those speaking a language (e.g., Russian), thereby showing that language barriers (e.g., for users that do not speak Russian) can create sub-communities with different knowledge and motivations. 
We further demonstrate how our method can identify the semantic meaning of {\it dark jargon} from its context, and discuss other potential applications of our approach. 
\end{abstract}

\begin{IEEEkeywords}
cybercrime, underground forum, dark keywords, slang, NLP, content analysis
\end{IEEEkeywords}

\section{Introduction}

Online underground forums are used for cyber-criminals to share knowledge, and trade illicit goods or services~\cite{caines2018automatically,motoyama2011analysis,pastrana2018crimebb, Bermudez2021shadyeconomy, bermudez2018under}. 
These forums provide unique opportunities to directly gather information about modern hacking tools and threats, allowing investigation of emerging cybercrime activities. 
Examining the content posted in these forums is of critical interest to the Cyber Threat Intelligence (CTI) industry, Law Enforcement Agencies (LEA), and criminologists.

Social media forensics dealing with underground forum data face important challenges~\cite{hughes2024art}. 
Accessing and collecting data from these forums requires dedicated crawling methods~\cite{ruiz2023general,turk2020tight,campobasso2019caronte}.
Also, data processing requires automatic tools and methods, due to the large volume of user-generated content~\cite{hughes2020detecting,portnoff2017tools, ramirez2021uncovering, arazzi2023nlp, torregrosa2023survey, rocha2017Authorship},
which is unstructured, noisy (including grammar or syntax errors), and has neologisms (dubbed {\it dark jargon}) in different languages ~\cite{Jin2022shredding,yuancantreader,Seyler2021darkjargon}. 

A crucial problem that has been overlooked is the analysis of the differences between communities that use multiple languages in forums. 
Geography is a known landmark for cybercrime hubs (popular ones being Russia, China, or Nigeria)~\cite{lusthaus2020mapping, edwards2018geography}, and language poses a natural barrier to accessing information that creates sub-communities of users with different literacy. 
By studying the differences in knowledge spread across languages, we can determine how knowledge is scattered, namely {\em pockets of knowledge}, allowing a fine-granular understanding of capabilities, the modeling of existing beliefs, and the profiling of actors discussing potential targets~\cite{greenberg2017entire}. 
Most existing works studying underground forums focus on one language, mainly English forums~\cite{Nunes2016darknetanddeepnet,caines2018automatically,pastrana2018crimebb,portnoff2017tools}. 
This limitation prevents the analysis of prominent underground forums and hinders the impact of their approach.
Prior work considering forum content in two languages (e.g., English and Russian) limits their focus to the aggregate of specific activities, i.e., without systematically bridging the semantic gap across topics and existing language barriers~\cite{Bhalerao2019mappingtheunderground}. 
More importantly, existing literature employs disparate approaches to study each language individually, without drawing connections to compare forum datasets across different languages~\cite{wagner2019cyber}. 

In this paper, we design methods and tools to study and investigate multilingual hacking forums. We examine the similarities and differences in knowledge exchanged between two sub-communities, using a unique dataset: a prominent multilingual invite-only hacking forum from which we crawled data using a custom stealth crawler.
The crawler has been operating for about 1.5 years, collecting  $\approx$18 years of historical data (totaling $\approx$1.1M posts and $\approx$156k conversations or threads).
We analyze the topics discussed and examine whether language creates a barrier that limits the information available to a monolingual reader, thereby determining if the language barrier is a driving factor that creates sub-communities even within the same hacking forum.
To extract meaningful and actionable information from this data, we combine state-of-the-art NLP techniques (sentence transformer) with an unsupervised clustering algorithm (HDBSCAN) to automatically structure the forum data. 
We then apply a post-processing step to represent the clustered topics through a customized application of Latent Dirichlet Allocation (LDA). These methods have proven effective for crime data analysis~\cite{ruellan2024conti, birks2020unsupervised}.
Using this approach, we automatically compare the content posted by different sub-communities, thereby revealing common topics as well as unique topics only discussed by one community, which we refer to as \textit{pockets of knowledge} (\pockets).
Finally, we provide case studies that demonstrate \pockets{} written in English and Russian, as well as show how our method can discover and understand dark keywords.

The main contributions of this paper are:
\begin{enumerate}
    \item We propose a flexible framework to automatically structure underground forum data and aggregate the contents discussed into common topics. 
    For each topic, we provide a semantic representation using a few keywords. 
    Due to the multilingual nature of the forum, we study different models and design the framework that does not require native-speaking understanding (\S\ref{sec:methodology}). 
    \item  We open the ground for a new direction toward studying underground forums by empirically testing different sentence embedding models. 
    We demonstrate how our method can compare multilingual content from a large criminal forum, identifying recurrent and novel topics that provide valuable insights to support online investigations (\S\ref{sec:analysis}). 
    \item We uncover \textit{Pockets of Knowledge}, and \textit{Dark Jargon} written in Russian, a challenge when understanding the behavior and language of cyber-criminals (\S\ref{sec:case-study}).
\end{enumerate}

\noindent Overall, our study allows for a better understanding, in an automatic way and for non-native analysts, of the topics being discussed in underground criminal forums. 
We discuss limitations, key findings, and potential real-world applications of our method in \S\ref{sec:discussion}. 
To foster research in the area and make our research reproducible, we publicly open-source our code in 
{{\url{https://github.com/mkmisch/pockets_of_knowledge}}}

\section{Limitations of Existing Approaches}

Comparing discussions of underground communities is inherently difficult due to the idiosyncrasies of the discussions and the lack of models trained with expert knowledge in the lingo used by hackers. 
The situation is further exacerbated by the multilingual nature of the hacking forum we are studying. Despite significant advances, criminologists often rely on manual analysis for reliable results~\cite{hutchings2019understanding,holt2022crime}. However, recent advances in text mining have made essential improvements for content analysis that assist in online investigations~\cite{portnoff2017tools,caines2018automatically, Gharibshah_Papalexakis_Faloutsos_2020, Stoddard_2021}, including the use of embeddings~\cite{Jin2022shredding} and clustering~\cite{Schafer2019}. 
Different from our methodology, these works target a unique language. 
Overall, we inherit, and partially tackle, existing challenges in the literature related to research in underground communities, such as data collection in adversarial environments~\cite{turk2020tight,campobasso2019caronte}, or the need for validation in the absence of ground truth, which often requires manual annotation~\cite{Li2021NEDetector,akyazi2021measuring} or the use of unsupervised methods~\cite{pastrana2018characterizing,Nunes2016darknetanddeepnet}. 
We refer the reader to a recent survey for a deeper understanding of the challenges ~\cite{hughes2024art}

Previous work targeted the detection and understanding of dark keywords, either using search engine optimization to derive new keywords~\cite{yangklingon}, or analyzing the semantics of words by comparing bad corpora with those from Reddit and Wikipedia~\cite{yuancantreader}. Other authors use Word Distribution Modeling and KL-Divergence, as well as cross-context lexical analysis~\cite{Seyler2021}, and in subsequent work use masked-language modeling for keyword detection, providing a lexicon for dark keywords~\cite{Seyler2021darkjargon}. NEDetector automatically identifies cybersecurity words by extracting word features and later applies Bidirectional LSTM~\cite{Li2021NEDetector}. 
Finally, recent work provides an analysis of emerging topics compared at different times, using a statistical approach on an English-speaking forum~\cite{hughes2020detecting}. 
Similar to our approach, Zhu et al. show that euphemisms used by fringe communities (such as hacking or criminal forums) can be detected by analyzing the surrounding contextual information~\cite{Zhu2021euphemism}. 
We, however, go beyond euphemism detection and characterize discussion topics across multiple languages.

Overall, we observe the following limitations in prior work:

\vspace{.1cm}
\noindent{\bf Scalability issues.}
The large amount of data requires automated processing to understand hacking content and extract topics~\cite{hughes2024art}. 
As such, many related works rely on manual efforts to label ground truth~\cite{Jin2022shredding, hutchings2019understanding, holt2022crime, caines2018automatically}, labels which are highly volatile due to the changing nature of the terminology used in the discussions~\cite{yuancantreader}.
We address scalability issues by using a combination of unsupervised methods to cluster semantically related conversational themes using sentence embeddings. 
Sentence embeddings are a powerful tool for dealing with unknown text as they effectively represent the overall meaning of a sentence rather than just individual words.

\vspace{.1cm}

\noindent{\bf Multilingual issues.}
Existing approaches such as \cite{rahman2021attackers,caines2018automatically, Schafer2019,yuancantreader} suffer limitations in the presence of multilingual neologisms. 
Other approaches to understanding forum content use features derived at the word or token level~\cite{pastrana2018characterizing}. 
This includes work focusing on the extraction of dark keywords \cite{Li2021NEDetector, yangklingon, Seyler2021, Seyler2021darkjargon, Zhao2016jargon, yuancantreader}.
However, unlike approaches based on sentence embeddings, word-based representations do not generalize well~\cite{reimers2019sentence}, neglecting the meaning of larger textual units. 
The presence of multilingual neologisms requires methods that can effectively comprehend the semantics of the conversation while dealing with unknown text. 
The context awareness of sentence embeddings enables handling ambiguity, understanding the relationships between the underlying intention and sentiment of text. At the same time, its ability to capture semantic similarities is essential for recognizing synonyms, addressing typographical errors, and identifying dark keywords.

\vspace{.1cm}
\noindent{\bf Gap:} There is a lack of multilingual analytical techniques capable of comprehensively capturing the complexity of underground forum discussions while dealing with jargon and neologisms typically seen in hacking forums.
We are the first to systematically compare topics in sub-communities of multilingual underground forums, addressing the lack of established baselines and the presence of dark keywords and noisy forum data.
Our approach presents a novel analytical technique of multilingual hacking forum data through the combination of multilingual handling, sentence embeddings, and clustering for data structuring, LDA keywords for topic representation, and keyword comparison for similarity measurement. 
We present a step-by-step validation, including manual approaches. 
We highlight that our method is data-driven, and the manual validation primarily serves for evaluation.


\section{Methodology}
\label{sec:methodology}
\begin{figure*}[t]
\centering
\includegraphics[width=0.9\textwidth, trim=70 52 100 280, clip]{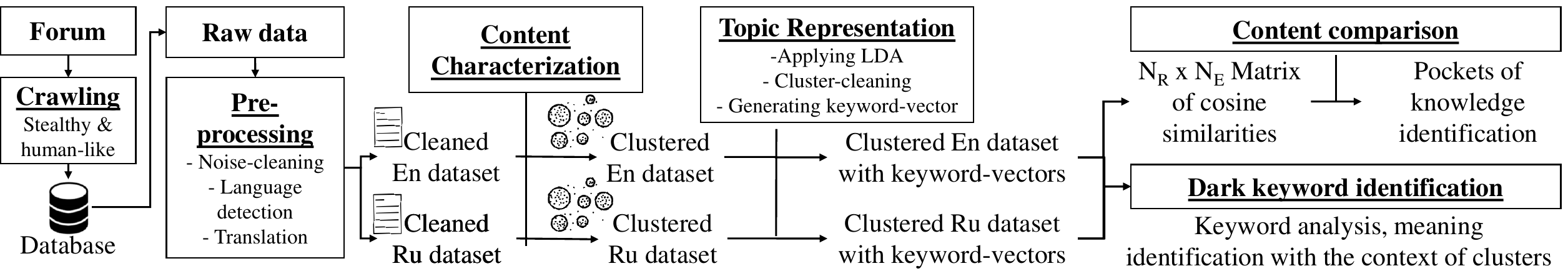} 
\caption{A graphical presentation of the methodology}
\label{fig:methodology}
\end{figure*}

Figure \ref{fig:methodology} depicts our methodology. We first collect forum data, which is pre-processed and split into two subsets (English and Russian), translating the Russian subset into English.
We characterize content using unsupervised clustering and apply topic modeling to each cluster. We then compare the similarities between English and Russian clusters and analyze common topics (discussed in both languages), and also \pockets{} (in a single language), as well as discover dark keywords and their hidden meaning.

\subsection{Data collection}
\label{subsec:crawling} 

Collecting data from underground communities requires dedicated methods~\cite{turk2020tight,campobasso2019caronte}.
A key feature of prominent underground forums is that they implement strict entry barriers~\cite{thomas2015framing, van2018plug}. 
Indeed, the hacking community we study is an invite-only forum dubbed as {\tt XIN}~\cite{mischinger2026xin}. 
Thus, covert crawling is critical due to the high cost of being detected (and banned).
We follow existing crawling approaches that mimic human navigation to fetch the entire forum content~\cite{campobasso2019caronte,pastrana2018crimebb}. 
We obtained Ethics clearance from our IRB (we provide Ethics discussion in ~\ref{sec:ethics}).

Our dataset, available through~\cite{mischinger2026xin}, comprises 53 sub-forums, with $\approx$1.1M posts and daily activity. According to the website, there are $\approx$85k registered members, whereas our dataset registers $\approx$34k members who posted at least once.
For our analysis, we consider 52 sub-forums (excluding a sub-forum that is only accessible to top-privileged members). 
The collected data spans over 18 years, starting from 2005 (the time of the first post seen in our subset). 
We focus our analysis on the title and the first post of every thread, since these are the most informative elements for the topic being discussed. In total, we have 156,348 threads started by $\approx$25k members. 
Most of the 52 sub-forums contain technical content, including hacking (e.g., malware, exploits, or social engineering attacks), cybercrime-related infrastructure (e.g., payment systems), and non-hacking content (e.g., web development or programming). Some non-technical sub-forums discuss job offers or auctions, entertainment (e.g., music or games), and general discussions regarding the forum.

\subsection{Data preprocessing}
\label{subsec:preprocessing}

The forum we study contains English and Russian texts.
Preliminary inspection shows that different languages often occur in separate paragraphs, so  we split posts using newlines as a delimiter. Then we use Google Translate for language detection and translation~\cite{googleTranslate}, as approaches based on automated machine translation have been demonstrated to be reliable for this task~\cite{mischinger2025lost}.
 
Besides regular conversations, forum data contains \othertext{}, e.g., code snippets, which we filter using two heuristics, i.e., it is mostly English and it includes a higher frequency of non-alphanumeric characters than in standard text. 
In English, there are on average 4\% of punctuations for every 5k characters~\cite{frequencyOfEnglishPunctuation,averageWordLength}.
However, forum data is messier, and a sentence might contain more punctuation (e.g., emoticons) and be shorter. 
We experimentally observe that a threshold of 12\% filters the highest amount of \othertext{}, whereby the percentage of actual conversational text falsely removed is less than 1\%. (in our validation set, only 0.4\% of FNs).
To prepare our dataset for sentence embedding, we remove very short text fragments, concretely, sentences with fewer than five spaces and 30 characters, and do a posterior manual check on the filtered data to verify that no relevant data is lost.
This removes irrelevant text such as empty strings or short tokens (e.g., \textit{good news:} or \textit{no-ip}).
Since the thread headlines typically are a representative description of the topic being discussed in the post, we keep this information even if the text is short. 
After pre-processing, we obtain 821,263 paragraphs across 176,033
threads, comprising 653,497 paragraphs in the Russian dataset (80\%) and 167,766 (20\%) in the English one.

\subsection{Content characterization}
\label{subsec:clustering}
We combine two well-known unsupervised methods: sentence embeddings for the feature extraction process, and clustering to group content based on semantic similarity. The former maps the paragraphs to multi-dimensional embedding vectors. The latter is used to group these vectors with HDBSCAN, a hierarchical density-based clustering algorithm \cite{Campello2013}.

We evaluate four different sentence embedding models, three monolingual (English generic~\cite{hugginface}, Russian generic~\cite{ruBert}, and English cybersecurity~\cite{secBERT}) and a multilingual~\cite{all-mpnet-multilingual,reimers2020making} (using the posts in their original languages). We select the model that works best on our multilingual underground forum content.

To determine the most suitable embedding model, we measure the performance of each model in combination with a density-based clustering approach. 
To avoid bias through fluctuation, we embed and cluster our dataset 5 times for each model, and take the average of the cluster outputs. We use three metrics: {\em Determinism} (i.e., consistency and stability of the output across five runs), using the ``RAND index''~\cite{rand1971objective}, which quantifies the consistency of sentence assignments across different clusterings (i.e., the number of sentences assigned to the same cluster set). Second, the amount of {\em outlier} data (i.e., data points that are not grouped into any cluster). Fewer outliers are desired to retain as much information as possible. Third, detected {\em variety in discussions} through the final distribution of clusters. 
A higher number of clusters indicates a greater topic diversity and better content differentiation. Using this, we observe that the monolingual English model. 

English and to-English-translated text outperforms the other models (see Table~\ref{tab:embedding-models}). 
Thus, we use this model for the rest of the analysis. 

The density-based clustering highly relies on the parameter \textit{minimum cluster size} (i.e., minimum number of data points to form a cluster).  
Assuming that different sub-forums do not necessarily have coinciding discussions, we select the highest minimum cluster size that yields meaningful clusters when clustering each sub-forum individually. 
We obtain a label for each cluster, consisting of up to 4 words: the most relevant verb, direct object, and two nouns.
These labels give an intuition of the cluster topics, but are not sufficient to draw conclusions about the actual content. Thus, we apply an extra topic representation approach, described in the following subsection.
Finally, to validate our clustering pipeline, we manually evaluate a subset of clusters to check if the clustered sentences belong to the same topic and are not falsely mixed together.

\begin{table}[htbp]
\caption{Comparison of sentence embedding models. }
\begin{center}
\begin{tabular}{|p{2.5cm}|c|c|c|}
\hline
\textbf{} & \textbf{Avg.} & \multicolumn{2}{c|}{\textbf{Avg. number of}}\\
\textbf{Model}& \textbf{{Rand index}} &  \multicolumn{1}{c}{\textbf{Outliers}}  & \textbf{Clusters}\\
\hline
\hline
{\bf English-generic}       &  {\bf 0.87} & \bf{450,816.8}    & {\bf 2,300.2} \\
\hline
Russian-generic       & 0.87        &  522,168.2	      & 816.2 \\
\hline
English-cybersec                 & 0.85        &  568,519.2	      & 1,868.8 \\
\hline
Multilingual & {0.88}        &  487,044.2	      & 2,049 \\
\hline
\end{tabular}
\label{tab:embedding-models}
\end{center}
\end{table}

\subsection{Topic representation}
\label{sec:topic_representation}
The clustering groups paragraphs by topic. 
The next step is to get  
a set of representative keywords to offer a better understanding of each topic.
To characterize the semantic content of each cluster, we leverage a well-known topic-modeling approach, i.e., LDA~\cite{Blei2003LatentDirichletAllocation}, to extract the most representative keywords for each cluster (topic). 
Concretely, we apply LDA to each cluster individually, setting the number of topics $\tau$ = 1 (since the data already belongs to the same topic). 
Additionally, we apply further steps to filter \othertext{}. First, we remove clusters where none of the four cluster labels' words are recognized by an `enriched' dictionary (containing all English words along with well-known IT words, like ``Skype'' or ``MySQL''). 
Second, we remove non-representative clusters based on the number of LDA keywords returned, specifically those where more than 80\% of the keywords are not recognized by our IT-words-enriched dictionary. 
We manually confirm that all the removed clusters only contain \othertext.
We then apply LDA to each cluster over both languages.

\subsection{Processing the representation}
\label{subsec:processing}
We use the representative keywords of the clusters for two different objectives.
First, to compare the two datasets (English and Russian) and discover {\it Pockets of Knowledge} (\pockets{}). Second, to uncover dark keywords and their hidden meaning.

\subsubsection{Comparing the content of the two datasets}
\label{subsubsec:comparing-content}
We compare all clusters of the English dataset with all clusters of the Russian dataset, through the cosine similarity ($s$) of their LDA keywords.
We characterize the similarity of two clusters using three levels: 1 (\textit{highly related}), 2  (\textit{somewhat related}), or 3 (\textit{not related}). 
{To find a threshold that informs similarity level, we follow a manual annotation and ranking approach, as done in previous work~\cite{Iqbal2023Nextdoor}.} Four annotators with expertise in underground forum analysis independently classify a subset of cluster pairs into one of the levels. Each human expert ranked the relatedness of two clusters by assigning a score level and discussing cases with clear disagreement. Afterward, we average the scores of the experts. 
We consider a topic as \commontopic{} if it is discussed in two \textit{highly related} clusters from both the English and Russian subsets, and as \uniquetopic, if it is present in a cluster for only one subset, meaning that all clusters are \textit{not related}, i.e., the topic is discussed only in either English or Russian. 

\subsubsection{Discovering dark keywords and their meaning}
Forum data contains forum slang and dark jargon, i.e., keywords whose usage or meaning is unknown in another context out of the underground community~\cite{Jin2022shredding,yuancantreader,Seyler2021darkjargon}. 
We use the comprehensive list of keywords generated for each cluster to identify such jargon. 
The list must be compared to known knowledge bases to exclude existing findings.
Recall that semantic clustering makes associations of sentences based on their context, and not just based on individual words. 
Thus, we analyze terms that appear as outliers within the context of their cluster.

\section{Analysis}
\label{sec:analysis}
This section first evaluates the algorithm for clustering paragraphs into topics.
Second, it describes the content analysis for detecting \pockets{} and understanding neologisms.

\subsection{Data clustering and evaluation}
\label{subsec:checking_if_model_works}
\begin{figure}[t]
\centering
\includegraphics[width=0.9\columnwidth]{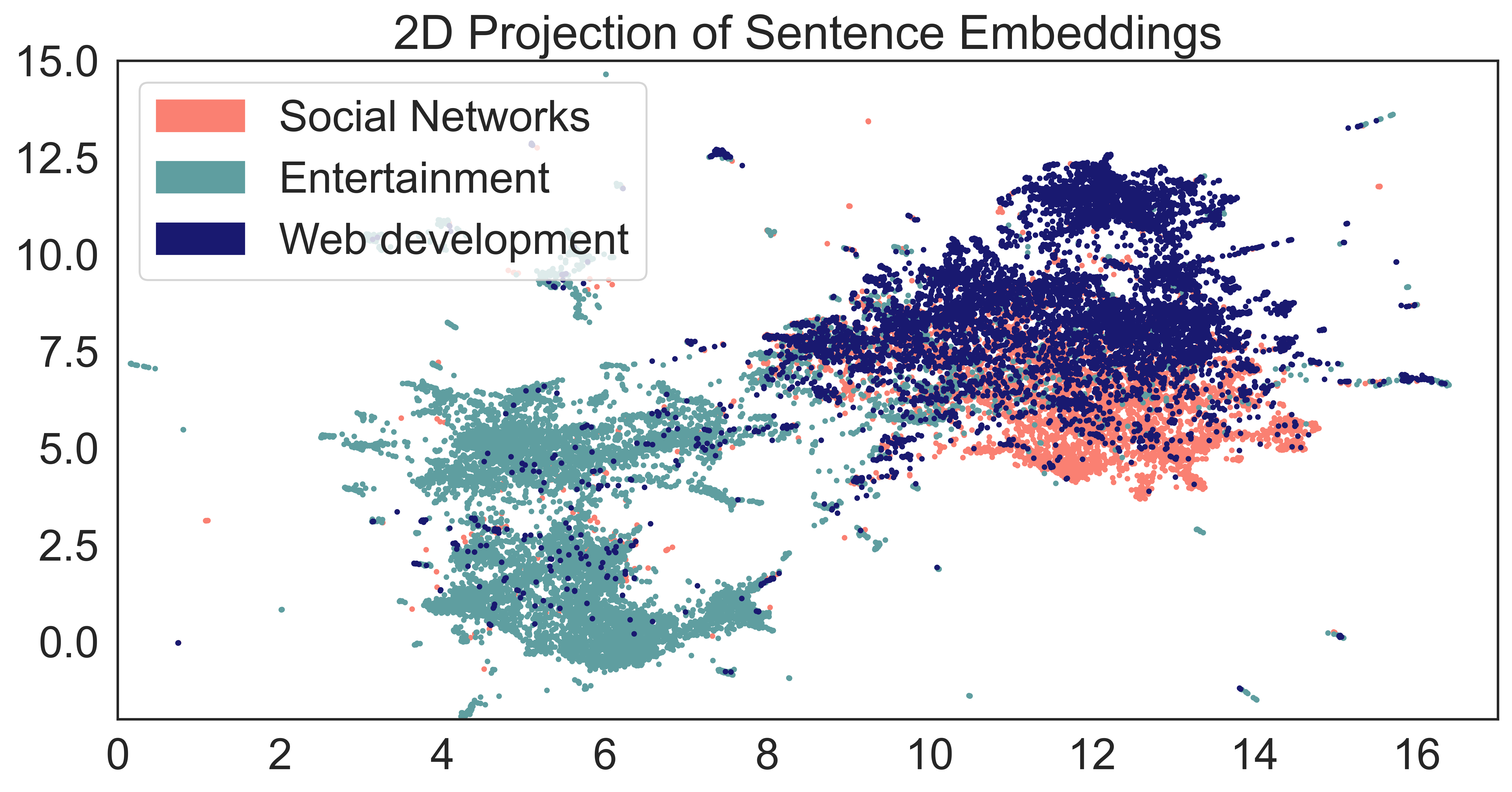}
\caption{Two-dimensional depiction of the vector embeddings for three different sub-forums.}
\label{fig:forums_4_21_58}
\end{figure}

To evaluate topic clustering, we generate sentence embeddings for three sub-forums with different content based on forum titles, and plot the two-dimensional representation to ease visualization in Figure \ref{fig:forums_4_21_58}. We observe clear distinctions between sub-forums related to communications, entertainment, and technical aspects. Still, we observe inter-forum similarities, which motivates the need to apply HDBSCAN to account for those similarities. 

After confirming that the proposed method successfully differentiates topics, we perform the clustering separately on the English and Russian datasets, using a \textit{minimum cluster size} of 24, obtaining 1,769 and 798 clusters respectively. 
To validate the model's reliability, we manually check a randomly chosen subset of 10\% of the clusters. 
We observe that the text within clusters is correctly clustered, with meaningful labels. 
Notably, our method relates synonyms and different words that convey similar meanings, even if the representative keyword is misspelled.
For example, it recognizes the word \textit{Zoom} as an application, or it clusters the words \textit{smb} and \textit{samba} together, as well as \textit{key logger} and \textit{keyboard log}.

\subsection{Topic representation}
We select 20 keywords per cluster. Fewer keywords lead to inaccuracy because a certain number of words is required to represent the content adequately; more keywords, however, lead to a higher amount of noise. 
After the LDA-filtering step, we remain with 1,667 Russian and 658 English clusters. After this final cleaning process, we study the number of 22,917 different authors and see that 6,975 write in both languages. This means that \textit{at most} 30\% of authors are bilingual (they post in the two languages). 

\subsubsection{Comparing the content}
\label{subsubsec:comparingthecontent} 
Computing the cosine similarity ($s$) of all 1,667 Russian clusters with all 658 English clusters leads to 1,096,886 similarity scores. 
We manually validate a random subset of 77 cluster pairs to obtain an appropriate threshold to detect whether two clusters are related to similar or distinct topics.
Consequently, for a given similarity $s_{ij}$ on clusters $C_i$ and $C_j$, we obtain the following threshold: 
    \begin{align}
     L(C_i,C_j) = \begin{dcases*}
         Highly\_Related, & if $ s_{ij} > 0.35 $,\\
         Somewhat\_Related, & if $ 0.35 \ge s_{ij} \ge 0.2 $ ,\\
         Not\_Related, & if $0.2 > s_{ij}$. 
         \end{dcases*}
    \end{align}
Table~\ref{tab:findings-cosine-sim} shows how related the clusters are and the expected cosine similarity of the LDA keywords. 
For each similarity score, we show the number of clusters in each dataset where this score is their maximum value.
For example, we observe 539 cluster pairs are \commontopics{}, but some appear repeated with other pairs (concretely, 342 English clusters, and 410 in the Russian dataset).

\begin{table}[htbp]
\centering
\caption{Computed cosine similarities ($s$) when comparing all Russian clusters with all English clusters.}
\begin{center}
\begin{tabular}{|c|c|c|c|c|}
\hline
 &  &  \textbf{Total} &  \multicolumn{2}{c|}{\textbf{\# Max. score}}\\
\cline{4-5} 
\textbf{Label} & \textbf{$s$} & \textbf{Pairs}  &  \textbf{Russian} & \textbf{English}\\
\hline
\hline
\textbf{Highly}    &    0.80 &                 1 &            1 &            1 \\
\textbf{Related}    &   0.75 &                 2 &            2 &            2 \\
    &   0.70 &                 2 &            2 &            2 \\
    &   0.65 &                 6 &            6 &            6 \\
    &   0.60 &                17 &           17 &           17 \\
    &   0.55 &                28 &           28 &           25 \\
    &   0.50 &                59 &           54 &           53 \\
    &   0.45 &               140 &          115 &           91 \\
    &   0.40 &               284 &          185 &          145 \\
\hline
\multicolumn{2}{|r|}{\textbf{TOTAL $>$ 0.35}} & \textbf{539} & \textbf{410} & \textbf{342}\\ 
\hline
\hline
\textbf{Somewhat} &       0.35 &               683 &          338 &          221 \\
\textbf{Related} &       0.30 &              1,601 &          561 &          305 \\
    &   0.25 &              4,447 &          938 &          421 \\
     &  0.20 &             12,619 &         1,308 &          514 \\
 \hline
\multicolumn{2}{|c|}{\textbf{TOTAL [0.35, 0.2]}} & \textbf{19,350} & \textbf{3,145} & \textbf{1,461}\\ 

\hline
\hline

\textbf{Not}  &       0.15 &             35,965 &         1574 &          593 \\
\textbf{Related}&       0.10 &            103,867 &         1658 &          650 \\
     &  0.05 &            280,230 &         1,667 &          658 \\
     &  0.00 &            656,935 &         1,667 &          658 \\
\hline
\multicolumn{2}{|c|}{\textbf{TOTAL $<$ 0.2}} & \textbf{107,6997} & \textbf{6,566} & \textbf{2,559}\\ 
\hline
\end{tabular}
\label{tab:findings-cosine-sim}
\end{center}
\end{table}

\vspace{.3cm}
\noindent\textbf{Common topics.}
The 539 \textit{highly related} cluster pairs in our dataset are \commontopics.
For reference, these pairs are listed in the Appendix we share in our repository \cite{appendix}. 
These topics span multiple conversations by 20,891 actors. 
Some clusters are paired more than once, i.e., they are paired together with different clusters that still talk about the same topic.
For example, an English cluster with label \texttt{sms\_sms\_bulk} discusses SMS (spam-) services and topics in general. 
Our method relates it to two Russian clusters, labeled as \texttt{sms\_sms\_number} and \texttt{sms\_spam} respectively. The former mostly describes steps to conduct bulk SMS spam. The latter discusses how to obtain and use services related to SMS (e.g., services used to deliver malicious SMS for malware delivery or phishing). 
While our semantic clustering does not group these topics together, they are highly related at different granular levels that are appropriate. 
This demonstrates how our methodology contextualizes topics within a forum, particularly highlighting the benefits of using LDA as a subsequent step to enhance the semantic analysis of the output from sentence embedding clustering.

\vspace{.3cm}
\noindent\textbf{Unique topics.} 
We found 306 clusters in the Russian dataset, and 127 clusters in the English dataset that are \uniquetopics{}.
 In both languages, we find \pockets{} from the 
 \textit{Entertainment}
 and \textit{Other} sub-forums. These mainly contain listings of, e.g., series or music bands, and are therefore very particular for a given sub-community. Because they are not hacking-related, we will not describe them further. 
We discuss Russian \pockets{} in \S\ref{sec:case-study}.

\subsubsection{Discovering neologisms and their meaning.}
\label{subsubsec:evaluation_dark_keywords}
Neologisms in underground communities play a key role, since these are terms used to describe novel cybercriminal activities, and are not necessarily part of any corpus of cybersecurity knowledge. 
As discussed, our methodology can be easily used to detect such neologisms. 
The LDA keywords (cf. \S\ref{sec:topic_representation}) allow for the detection of such neologisms. 
For each cluster, we filter out terms from our knowledge base, i.e., the DarkJargon~\cite{Seyler2021darkjargon} and the NCSC~\cite{ncscglossary} glossaries. 
This way, we reduce the total number of keywords from  {8,101 to 8,034.}
This final set also contains words from the regular English vocabulary. 
We do not remove these, since neologisms might be homonyms, e.g., ``rat'', as a Remote Access Trojan.
In \S\ref{subsec:case_study_darkKeywords} we study each suspicious term within the context of its cluster.

\section{Case Study}
\label{sec:case-study}
Leveraging our methodology, we describe two different case studies. First, we investigate \pockets{} and discuss the threat they pose. 
Second, we demonstrate how our approach uncovers CTI-relevant keywords and their meaning.

\subsection{Pockets of knowledge}
\label{subsec:pockets}
In the 306 Russian \uniquetopics{}, we observe various technical (hacking) discussions, as well as social engineering-related topics (intended for psychological and social manipulation), which offer a unique understanding of current and prospective social engineering methods that put those with access to these \pockets{} in a vantage point in the arms race.

\noindent\textbf{Technical, hacking-related topics. }
One actor shared detailed instructions for hacking the Basic Input/Output System (BIOS) at boot time. 
In particular, the post describes how to reset the BIOS password and conduct brute-force attacks using universal passwords. 
Furthermore, the actor provides instructions to manipulate the motherboard hardware to reset BIOS settings, as well as a tutorial including example code on how to use software to destroy the motherboard memory content, which would then lead to a password reset. 
This is a serious threat since compromising the BIOS is one of the first steps to install advanced malware such as Rootkits~\cite{hili2014bios}, often attributed to high-profile actors.
In another example, the community seeks and shares advice for securing and breaking physical locks. 
Finally, one thread discusses the blockchain ``double-spending'' attack, showcasing how one could theoretically gain over 50\% of network power over different periods, and listing existing double-spending attacks.

\noindent\textbf{Interaction with Law Enforcement Agencies. }
We found authors discussing potential issues and ways to deal with LEA officers. 
First, they discuss the threats of computer and network logs to their criminal endeavors.   
Concretely, authors comment on Russian laws, arguing that log files are valid and recurrent forensic evidence, and advocating for their removal to hinder prosecution. 
Additionally, they provide tips on how to act with the (Russian) police in terms of rights and duties. 
They discuss manipulation tricks that police might use, e.g., during an interrogation or home raids, and how to manage emotions that could make the actor suspicious, like fright, intimidation, and humiliation. 
These discussions provide Russian speakers with new insights on how to circumvent prosecution by LEA officers.

\noindent\textbf{Social manipulation. }
We found users sharing know-how on how to ``manipulate'' and ``read'' people, including a detailed step-by-step guide to develop skills to make people uncomfortable, thus getting interpersonal advantage when interacting. 
There are also discussions about ``women's signals''. Women would giggle and steal glances when they liked a guy, look down and away with feigned modesty, and show their legs by lifting the edge of their skirt to attract men.
Finally, a thread describes the philosophy of lies in Russian culture, the motivation behind lies, and how to detect them.

\subsubsection{Comparison with English}
We systematically analyze the English dataset to confirm whether the identified Russian cases are \pockets{}. 
Concretely, we look for coincidences of the Russian LDA keywords on the English clusters. When matching, we analyze the cluster content and the entire threads using thematic analysis \cite{lochmiller2021conducting}.
We found a conversation in the English dataset related to the BIOS hacking topic, but it only asked for help, rather than providing information. 
Besides that, there are no other discussions on bios hacking.
Apart from this, we find no other English threads related to the topics described previously, and thus we confirm that these are Russian \pockets.

\subsection{Understanding dark keywords} 
\label{subsec:case_study_darkKeywords}
To uncover dark keywords, we inspect the list of 8k words (cf. \S\ref{subsubsec:evaluation_dark_keywords}). We found various interesting neologisms, shown in Table~\ref{tab:dark-keywords}, {together with the context taken from the cluster, and other associated LDA keywords}. 
A description of these neologisms is finally obtained from various public Open Source Intelligence (OSINT) sources. 
We observe how our method associates terms that were previously unknown in our knowledge base, to criminal activities such as network attacks (\textit{hvnc}), malware families (\textit{lockbit}), botnets (\textit{rustock}), or money laundering services (\textit{tumblebit}). 
Also, we note the usefulness of the context provided by the LDA keywords. For example, \textit{hvnc} is associated with terms  {\it connect}, {\it desktop} and {\it rdp} (Remote Desktop protocol), or \textit{bootkit}, to {\it malicious}, {\it control}, {\it boot} and {\it sector}.
\begin{table*}[htbp]
\caption{Findings of dark keywords. }
\begin{center}
\begin{tabular}{|c|p{4cm}|p{4cm}|p{4cm}|}
\hline
\textbf{Neologism} & \textbf{LDA keywords (selection)} & \textbf{Meaning from cluster context}   & \textbf{Validation (OSINT)}  \\
\hline
\hline
bootkit	&	boot, control, disk, hard, malicious, sector	&	A malware targeting the early stages of system boot &	A malware targeting the boot sectors. 	\\
\hline		
clickbot &	bot, click, clicker fraud, software	&	Fraud software that automatically clicks links or similar	&	Software for click-fraud\\
\hline							
hvnc	&	connect, desktop, login, rdp, rat, botnet	& A connection function included in some remote access trojans (RAT)	&	Module which allows attackers to gain user-grade access to an infected PC	\\
\hline							
lockbit	&	encryption, malicious, attack, ransomhack	&	A ransomware	&	A ransomware	\\
\hline							
rustock	&	code, computer, infected, program, rootkit	&	A botnet, using rootkit functionality for spam mailing	&	A botnet	\\
\hline							
tdss	&	affiliate, botnet, malware, program, rootkit	&	Rootkit, distribution of malware through affiliate program	&	A trojan and rootkit \\
\hline
\end{tabular}
\label{tab:dark-keywords}
\end{center}
\end{table*}

\section{Discussion}
\label{sec:discussion}
This section describes the limitations, the most important findings, and the potential implications of the study presented. 

\subsection{Limitations}
\label{subsec:limitations}

\noindent\textbf{Dataset.}
Our evaluation focuses on thread headlines and first posts.  There may be more extensive knowledge in unexplored parts of the forum. 
However, we consider this unlikely, since we capture the main elements of a thread, and post replies usually relate to the topic of headlines and first posts. 
It is also possible that for multiple runs with different seeds, the clustering and LDA keywords might observe different configurations.
This may lead the algorithm to group \commontopics{} or \uniquetopics{} differently. 
To mitigate this, we re-run our experiments multiple (\textgreater 10) times, observing consistent results across runs. 
Finally, we limit our analysis to a single forum due to resource constraints. 
Despite this, we show how our method detects \pockets{} as discussed above. 
We conclude that one single forum suffices to answer our RQs, generalizing across the data we observe. 
We plan to study the prevalence of \pockets{} across forums as part of our future work.

\vspace{0.2cm}

\noindent\textbf{Text clutter and typos.}
Our data cleaning process aims to remove as much \othertext{} as possible; however, this step is not perfect.  
Interestingly, we observe that such \othertext{} forms its own cluster, enabling its distinction from actual conversations. 
Furthermore, we found various instances of such content during our manual validation, mainly in the \uniquetopics{} from the English. While it requires further efforts to improve the data cleaning step, we believe that our method successfully processes unstructured, noisy forum data. 
One limitation observed is that different terms naming the same thing, or syntax errors occurring in these, are not recognized, e.g., \textit{keilogger} and \textit{keylogger}. Thus, a follow-up step is to combine LDA with word embeddings. A challenge to be addressed is that generating precise LDA keywords requires clusters with a sufficient amount of text, which is not always the case.

\vspace{0.2cm}
\noindent\textbf{Newline splitting.}
We split posts by newlines to separate content written in different languages. 
While we found this separation ideal for our dataset, in other datasets, a different delimiter may be necessary. 
We note that there is no perfect cut when splitting sentences in unstructured text~\cite{Li2022unstructured, hughes2024art}.
While newlines generally work well in our domain, we encounter cases where the splitting method separates text that is part of the same sentence, for example, in cases where the text appears to be copy-pasted (such as a shared article).
Despite this, we have found that the discussion topic is still accurately represented when fractions of the entire text are clustered correctly.

\vspace{0.2cm}
\noindent\textbf{Machine translation.}
Using machine translation might lead to translation inaccuracies in the presence of dark keywords and forum slang. 
However, recent work showed that machine translation applied to underground forums works reliably~\cite{mischinger2025lost}. We furthermore analyzed different sentence-embedding models, in \ref{subsec:clustering}. Our analysis concluded that our current pipeline (with machine translation) performs best.

\subsection{Key Findings}
Our method provides easy-to-understand models for analysis, accessible for non-native speakers and LEA officers or researchers with little background on hacking-related jargon, that offer substantial benefits over the limitations discussed above.

\subsubsection{Understanding topics in the absence of ground truth} 
\label{subsecsec:key-findings_absence_of_ground_truth}
Our method systematically processes noisy and unstructured data to gain a better understanding of the topics discussed in underground criminal forums. 
Combining paragraph embeddings with a clustering algorithm proves beneficial for non-native investigators. 
We also see that splitting text by newlines amplifies context.
In underground forums, this is especially useful due to the particularities of the posts, which include dark jargon, neologisms, and several grammar and syntax errors. 
Thus, our work suggests that having a greater context has the potential to offer significant new insights into the discovery of threats.

\subsubsection{Understanding and comparing sub-communities} 
The application of LDA allows us to compare the content of the clusters from different sub-communities. 
To the best of our knowledge, we are the first to investigate these differences in underground forums.
In our work, the data we compare stems from the same forum and may have overlapping topics, as expected. Despite this, we can show differences in knowledge across the two sub-communities we analyze. 
We believe our method can be applied to compare the content of other forums/\-communities to uncover differences in knowledge, as we will explore in future work.
However, we argue that our method is equally valuable when comparing two datasets where scarcely related topics are expected. 
The number of unrelated \uniquetopics{} will be unsurprisingly high, while \commontopics{} will uncover hidden related discussions and common trends.

\section{Implications and Conclusion}
{In this work, we investigated a multilingual English-Russian underground forum and proposed a methodology that systematically untangles and compares topics between hacking sub-communities, facing the challenge of processing a large amount of noisy and unstructured data, containing slang, shortcuts, and dark keywords, in the absence of ground truth.}
We have provided empirical evidence of the proliferation of knowledge around various social engineering campaigns that are available only to those speaking Russian. 
Furthermore, we showed that our approach can be used to detect dark jargon and uncover its hidden meaning. 
Our integrated approach allows for deeper insights into the structure and dynamics of these forums, and we identify two main applications where our work has the potential for significant impact.

\vspace{.1cm}
\noindent{ \bf Cybercrime and Cyber Threat Intelligence.} 
Our case studies demonstrated how unique insights into hacking sub-communities can improve targeted CTI efforts. 
Moreover, by not only identifying but also contextualizing dark keywords, our approach enables CTI to uncover unknown threats beyond existing glossaries. 
We systematize critical processes in the analysis of cyber threats that can dramatically reduce operational costs.
Existing methods suffer limitations in the presence of multilingual neologisms \cite{rahman2021attackers,caines2018automatically}, or rely on manual annotation for ground truth~\cite{Jin2022shredding, hutchings2019understanding, holt2022crime, caines2018automatically}.
Our method can be a cornerstone to address these issues as it combines modern NLP techniques designed to address the challenge of studying cyber-related multilingual data containing dark keywords. 
Whereas our method has been applied to a static dataset (i.e., collected and stored offline), it could be deployed in real-time threat monitoring. 
Indeed, NLP-driven analysis of multilingual underground forums has already proven valuable to predict actionable CTI using only the posts' context and content~\cite{mischinger2024IOCs}.

\vspace{.1cm}
\noindent{\bf Early detection of deviant behavior.}
Our approach successfully groups topics into clusters and provides a list of relevant keywords describing each cluster. These keywords provide a quick overview of the content to cybersecurity analysts, enabling the detection of outlier topics --- discussions that are not expected to be there. 
A direct application of this is the detection of conversations in forums connected to terrorism or extremism, which have been reported to occur in general-purpose forums~\cite{denning2013terror,gaudette2022role}. 
In the case discussed by Gaudette et al.~\cite{gaudette2022role}, a forum focused on music was used by far-right extremists to recruit new members, by encouraging users to attend concerts from white-power bands. Our method can aid in the prompt identification of such conversations in the same way we identify pockets of knowledge in hacking forums.

\section{Ethics}
\label{sec:ethics}
This research has some Ethical implications. First, there is a risk to the privacy of the users of the forum. We make no attempt at deanonymization in cases where the users use pseudonyms, and we do not focus on the real identities of the people involved in cases where the users use their legal names. 
Second, to prevent disruption of the normal operation of the site, our crawler mimics human behavior. 
Finally, we avoid sharing any names or the forum’s identity to avoid promotion and minimize cybercrime risks.
We have discussed these risks with the Institutional Review Board and obtained approval to collect data and perform our analysis.

\section{Acknowledgements}
This project was primarily funded by TED2021-132900A-I00,
and TED2021-132170A-I00,
from the Spanish Ministry of Science and Innovation, funded by MCIN/AEI\-/10.13039\-/501100011033, and the European Union-NextGenerationEU/PRTR, 
as well as PID2022-143304OB-I00,
funded by MCIN/AEI\-/10.13039/\-501100011033/ and the ERDF ''A way of making Europe.''
Guillermo Suárez Tangil is a 2020 RyC fellow RYC2020-029401-I funded by MCIU/AEI/10.13039/501100011033 and the ESF Investing in your future. The same grant has funded Mariella Mischinger's work.

\bibliographystyle{IEEEtran}

\bibliography{reference_LONG_VERSION}

@inproceedings{mischinger2026xin,
  title={Crime VIP: A Closed-Access Underground Hacking Forum},
  author={Mischinger, Mariella and Pastrana, Sergio and Suarez-Tangil, Guillermo},
  booktitle={Proceedings of the International Conference on Web and Social Media (ICWSM)},
  year={2026},
  organization={AAAI Press}
}

@inproceedings{mischinger2024IOCs,
  title={IoC Stalker: Early detection of Indicators of Compromise},
  author={Mischinger, Mariella and Pastrana, Sergio and Suarez-Tangil, Guillermo},
  booktitle={Proceedings of the 40th Annual Computer Security Applications Conference},
  year={2024}
}

@inproceedings{mischinger2025lost,
  title={Lost in Translation: Analyzing Non-English Cybercrime Forums},
  author={Mischinger, Mariella and Hughes, Jack and Vitiugin, Fedor and  Pastrana, Sergio and Hutchings, Alice and Suarez-Tangil, Guillermo},
  booktitle={2025 APWG Symposium on Electronic Crime Research (eCrime)},
  year={2025},
  organization={IEEE}
}

@INPROCEEDINGS{Zhu2021euphemism,
  author={Zhu, Wanzheng and Gong, Hongyu and Bansal, Rohan and Weinberg, Zachary and Christin, Nicolas and Fanti, Giulia and Bhat, Suma},
  booktitle={2021 IEEE Symposium on Security and Privacy (SP)}, 
  title={Self-Supervised Euphemism Detection and Identification for Content Moderation}, 
  year={2021},
  volume={},
  number={},
  pages={229-246},
  doi={10.1109/SP40001.2021.00075}}

@inproceedings{akyazi2021measuring,
  title={Measuring cybercrime as a service (caas) offerings in a cybercrime forum},
  author={Akyazi, Ugur and van Eeten, MJG and Ganan, C Hernandez},
  booktitle={Workshop on the Economics of Information Security},
  year={2021}
}

@article{hughes2024art,
  title={The art of cybercrime community research},
  author={Hughes, Jack and Pastrana, Sergio and Hutchings, Alice and Afroz, Sadia and Samtani, Sagar and Li, Weifeng and Santana Marin, Ericsson},
  journal={ACM Computing Surveys},
  volume={56},
  number={6},
  pages={1--26},
  year={2024},
  publisher={ACM New York, NY}
}

@article{caines2018automatically,
  title={Automatically identifying the function and intent of posts in underground forums},
  author={Caines, Andrew and Pastrana, Sergio and Hutchings, Alice and Buttery, Paula J},
  journal={Crime Science},
  volume={7},
  number={1},
  pages={1--14},
  year={2018},
  publisher={Springer}
}

@inproceedings{hughes2020detecting,
  title={Detecting Trending Terms in Cybersecurity Forum Discussions},
  author={Hughes, Jack and Aycock, Seth and Caines, Andrew and Buttery, Paula and Hutchings, Alice},
  booktitle={Proceedings of the Sixth Workshop on Noisy User-generated Text (W-NUT 2020)},
  organization={Association for Computational Linguistics},
  pages={107--115},
  year={2020}
}

@inproceedings{pastrana2018characterizing,
  title={Characterizing eve: Analysing cybercrime actors in a large underground forum},
  author={Pastrana, Sergio and Hutchings, Alice and Caines, Andrew and Buttery, Paula},
  booktitle={Research in Attacks, Intrusions, and Defenses: 21st International Symposium, RAID 2018, Heraklion, Crete, Greece, September 10-12, 2018, Proceedings 21},
  pages={207--227},
  year={2018},
  organization={Springer}
}

@article{ruiz2023general,
  title={A general and modular framework for dark web analysis},
  author={Ruiz R{\'o}denas, Jos{\'e} Manuel and Pastor-Galindo, Javier and G{\'o}mez M{\'a}rmol, F{\'e}lix},
  journal={Cluster Computing},
  pages={1--17},
  year={2023},
  publisher={Springer}
}

@inproceedings{reimers2019sentence,
    title = "Sentence-{BERT}: Sentence Embeddings using {S}iamese {BERT}-Networks",
    author = "Reimers, Nils  and
      Gurevych, Iryna",
    editor = "Inui, Kentaro  and
      Jiang, Jing  and
      Ng, Vincent  and
      Wan, Xiaojun",
    booktitle = "Proceedings of the 2019 Conference on Empirical Methods in Natural Language Processing and the 9th International Joint Conference on Natural Language Processing (EMNLP-IJCNLP)",
    month = nov,
    year = "2019",
    address = "Hong Kong, China",
    publisher = "Association for Computational Linguistics",
    url = "https://aclanthology.org/D19-1410",
    doi = "10.18653/v1/D19-1410",
    pages = "3982--3992",
}

@article{torregrosa2023survey,
  title={A survey on extremism analysis using natural language processing: definitions, literature review, trends and challenges},
  author={Torregrosa, Javier and Bello-Orgaz, Gema and Mart{\'\i}nez-C{\'a}mara, Eugenio and Ser, Javier Del and Camacho, David},
  journal={Journal of Ambient Intelligence and Humanized Computing},
  volume={14},
  number={8},
  pages={9869--9905},
  year={2023},
  publisher={Springer}
}

@article{ramirez2021uncovering,
  title={Uncovering cybercrimes in social media through natural language processing},
  author={Ram{\'\i}rez S{\'a}nchez, Juli{\'a}n and Campo-Archbold, Alejandra and Zapata Rozo, Andr{\'e}s and D{\'\i}az-L{\'o}pez, Daniel and Pastor-Galindo, Javier and G{\'o}mez M{\'a}rmol, F{\'e}lix and Aponte D{\'\i}az, Juli{\'a}n},
  journal={Complexity},
  volume={2021},
  pages={1--15},
  year={2021},
  publisher={Hindawi Limited}
}

@article{arazzi2023nlp,
  title={NLP-Based Techniques for Cyber Threat Intelligence},
  author={Arazzi, Marco and Arikkat, Dincy R and Nicolazzo, Serena and Nocera, Antonino and Conti, Mauro and others},
  journal={arXiv preprint arXiv:2311.08807},
  year={2023}
}

@ARTICLE{rocha2017Authorship,
  author={Rocha, Anderson and Scheirer, Walter J. and Forstall, Christopher W. and Cavalcante, Thiago and Theophilo, Antonio and Shen, Bingyu and Carvalho, Ariadne R. B. and Stamatatos, Efstathios},
  journal={IEEE Transactions on Information Forensics and Security}, 
  title={Authorship Attribution for Social Media Forensics}, 
  year={2017},
  volume={12},
  number={1},
  pages={5-33},
  doi={10.1109/TIFS.2016.2603960}}

@article{Gharibshah_Papalexakis_Faloutsos_2020, title={REST: A Thread Embedding Approach for Identifying and Classifying User-Specified Information in Security Forums}, volume={14}, url={https://ojs.aaai.org/index.php/ICWSM/article/view/7293}, DOI={10.1609/icwsm.v14i1.7293}, abstractNote={&lt;p&gt;How can we extract useful information from a security forum? We focus on identifying threads of interest to a security professional: (a) alerts of worrisome events, such as attacks, (b) offering of malicious services and products, (c) hacking information to perform malicious acts, and (d) useful security-related experiences. The analysis of security forums is in its infancy despite several promising recent works. Novel approaches are needed to address the challenges in this domain: (a) the difficulty in specifying the “topics” of interest efficiently, and (b) the unstructured and informal nature of the text. We propose, REST, a systematic methodology to: (a) identify threads of interest based on a, possibly incomplete, bag of words, and (b) classify them into one of the four classes above. The key novelty of the work is a multi-step weighted embedding approach: we project words, threads and classes in appropriate embedding spaces and establish relevance and similarity there. We evaluate our method with real data from three security forums with a total of 164k posts and 21K threads. First, REST robustness to initial keyword selection can extend the user-provided keyword set and thus, it can recover from missing keywords. Second, REST categorizes the threads into the classes of interest with superior accuracy compared to five other methods: REST exhibits an accuracy between 63.3-76.9%. We see our approach as a first step for harnessing the wealth of information of online forums in a user-friendly way, since the user can loosely specify her keywords of interest.&lt;/p&gt;}, number={1}, journal={Proceedings of the International AAAI Conference on Web and Social Media}, author={Gharibshah, Joobin and Papalexakis, Evangelos E. and Faloutsos, Michalis}, year={2020}, month={May}, pages={217-228} }

@article{Stoddard_2021, title={Popularity Dynamics and Intrinsic Quality in Reddit and Hacker News}, volume={9}, url={https://ojs.aaai.org/index.php/ICWSM/article/view/14636}, DOI={10.1609/icwsm.v9i1.14636}, abstractNote={ &lt;p&gt; In this paper we seek to understand the relationship between the online popularity of an article and its intrinsic quality. Prior experimental work suggests that the relationship between quality and popularity can be very distorted due to factors like social influence bias and inequality in visibility. We conduct a study of popularity on two different social news aggregators, Reddit and Hacker News. We define quality as the number of votes an article would have received if each article was shown, in a bias-free way, to an equal number of users. We propose a simple Poisson regression method to estimate this quality metric from time-series voting data. We validate our methods on data from Reddit and Hacker News, as well the experimental data from prior work. Using these estimates, we find that popularity on Reddit and Hacker News is a relatively strong reflection of intrinsic quality. &lt;/p&gt; }, number={1}, journal={Proceedings of the International AAAI Conference on Web and Social Media}, author={Stoddard, Greg}, year={2021}, month={Aug.}, pages={416-425} }

@inproceedings{hutchings2019understanding,
  title={Understanding ewhoring},
  author={Hutchings, Alice and Pastrana, Sergio},
  booktitle={2019 IEEE European Symposium on Security and Privacy (EuroS\&P)},
  pages={201--214},
  year={2019},
  organization={IEEE}
}

@article{holt2022crime,
  title={A crime script model of Dark web Firearms Purchasing},
  author={Holt, Thomas J and Lee, Jin Ree},
  journal={American Journal of Criminal Justice},
  pages={1--21},
  year={2022},
  publisher={Springer}
}

@inproceedings{motoyama2011analysis,
  title={An analysis of underground forums},
  author={Motoyama, Marti and McCoy, Damon and Levchenko, Kirill and Savage, Stefan and Voelker, Geoffrey M},
  booktitle={Proceedings of the 2011 ACM SIGCOMM conference on Internet measurement conference},
  pages={71--80},
  year={2011}
}

@article{greenberg2017entire,
  title={How an entire nation became Russia’s test lab for cyberwar},
  author={Greenberg, Andy},
  journal={Wired, June},
  volume={20},
  year={2017}
}

@article{gaudette2022role,
  title={The role of the internet in facilitating violent extremism: insights from former right-wing extremists},
  author={Gaudette, Tiana and Scrivens, Ryan and Venkatesh, Vivek},
  journal={Terrorism and Political Violence},
  volume={34},
  number={7},
  pages={1339--1356},
  year={2022},
  publisher={Taylor \& Francis}
}

@incollection{denning2013terror,
  title={Terror’s web: how the internet is transforming terrorism},
  author={Denning, Dorothy E},
  booktitle={Handbook of internet crime},
  pages={212--231},
  year={2013},
  publisher={Willan}
}

@inproceedings{portnoff2017tools,
  title={Tools for automated analysis of cybercriminal markets},
  author={Portnoff, Rebecca S and Afroz, Sadia and Durrett, Greg and Kummerfeld, Jonathan K and Berg-Kirkpatrick, Taylor and McCoy, Damon and Levchenko, Kirill and Paxson, Vern},
  booktitle={Proceedings of the 26th international conference on world wide web},
  pages={657--666},
  year={2017}
}

@inproceedings{pastrana2018crimebb,
  title={Crimebb: Enabling cybercrime research on underground forums at scale},
  author={Pastrana, Sergio and Thomas, Daniel R and Hutchings, Alice and Clayton, Richard},
  booktitle={Proceedings of the 2018 World Wide Web Conference},
  pages={1845--1854},
  year={2018}
}

@inproceedings{campobasso2019caronte,
  title={Caronte: crawling adversarial resources over non-trusted, high-profile environments},
  author={Campobasso, Michele and Burda, Pavlo and Allodi, Luca},
  booktitle={2019 IEEE European Symposium on Security and Privacy Workshops (EuroS\&PW)},
  pages={433--442},
  year={2019},
  organization={IEEE}
}

@InProceedings{Campello2013,
author="Campello, Ricardo J. G. B.
and Moulavi, Davoud
and Sander, Joerg",
editor="Pei, Jian
and Tseng, Vincent S.
and Cao, Longbing
and Motoda, Hiroshi
and Xu, Guandong",
title="Density-Based Clustering Based on Hierarchical Density Estimates",
booktitle="Advances in Knowledge Discovery and Data Mining",
year="2013",
publisher="Springer Berlin Heidelberg",
address="Berlin, Heidelberg",
pages="160--172",
isbn="978-3-642-37456-2"
}

@inproceedings{turk2020tight,
  title={A tight scrape: Methodological approaches to cybercrime research data collection in adversarial environments},
  author={Turk, Kieron and Pastrana, Sergio and Collier, Ben},
  booktitle={2020 IEEE European Symposium on Security and Privacy Workshops (EuroS\&PW)},
  pages={428--437},
  year={2020},
  organization={IEEE}
}

@article{Schafer2019,
   author = {Matthias Schafer and Markus Fuchs and Martin Strohmeier and Markus Engel and Marc Liechti and Vincent Lenders},
   doi = {10.23919/CYCON.2019.8756845},
   isbn = {9789949990443},
   issn = {23255374},
   journal = {International Conference on Cyber Conflict, CYCON},
   month = {5},
   publisher = {NATO CCD COE Publications},
   title = {BlackWidow: Monitoring the Dark Web for Cyber Security Information},
   volume = {2019-May},
   year = {2019},
}

@inproceedings{Seyler2021,
  title={Towards dark jargon interpretation in underground forums},
  author={Seyler, Dominic and Liu, Wei and Wang, XiaoFeng and Zhai, ChengXiang},
  booktitle={European Conference on Information Retrieval},
  pages={393--400},
  year={2021},
  organization={Springer}
}

@inproceedings{Seyler2021darkjargon,
  title={Darkjargon. net: A platform for understanding underground conversation with latent meaning},
  author={Seyler, Dominic and Liu, Wei and Zhang, Yunan and Wang, XiaoFeng and Zhai, ChengXiang},
  booktitle={Proceedings of the 44th International ACM SIGIR Conference on Research and Development in Information Retrieval},
  pages={2526--2530},
  year={2021}
}

@inproceedings{yuancantreader,
  title={Reading thieves' cant: automatically identifying and understanding dark jargons from cybercrime marketplaces},
  author={Yuan, Kan and Lu, Haoran and Liao, Xiaojing and Wang, XiaoFeng},
  booktitle={27th USENIX Security Symposium (USENIX Security 18)},
  pages={1027--1041},
  year={2018}
}

@inproceedings{yangklingon,
  title={How to learn klingon without a dictionary: Detection and measurement of black keywords used by the underground economy},
  author={Yang, Hao and Ma, Xiulin and Du, Kun and Li, Zhou and Duan, Haixin and Su, Xiaodong and Liu, Guang and Geng, Zhifeng and Wu, Jianping},
  booktitle={2017 IEEE Symposium on Security and Privacy (SP)},
  pages={751--769},
  year={2017},
  organization={IEEE}
}

@article{Li2021NEDetector,
  title={NEDetector: Automatically extracting cybersecurity neologisms from hacker forums},
  author={Li, Ying and Cheng, Jiaxing and Huang, Cheng and Chen, Zhouguo and Niu, Weina},
  journal={Journal of Information Security and Applications},
  volume={58},
  pages={102784},
  year={2021},
  publisher={Elsevier}
}

@inproceedings{thomas2015framing,
  title={Framing dependencies introduced by underground commoditization},
  author={Thomas, Kurt and Huang, Danny and Wang, David and Bursztein, Elie and Grier, Chris and Holt, Thomas J and Kruegel, Christopher and McCoy, Damon and Savage, Stefan and Vigna, Giovanni},
  booktitle = {Annual Workshop on the Economics of Information Security (WEIS)},
  year={2015}
}

@inproceedings{van2018plug,
  title={Plug and prey? measuring the commoditization of cybercrime via online anonymous markets},
  author={Van Wegberg, Rolf and Tajalizadehkhoob, Samaneh and Soska, Kyle and Akyazi, Ugur and Ganan, Carlos Hernandez and Klievink, Bram and Christin, Nicolas and Van Eeten, Michel},
  booktitle={27th USENIX security symposium (USENIX security 18)},
  pages={1009--1026},
  year={2018}
}

@article{Jin2022shredding,
   author = {Youngjin Jin and Eugene Jang and Yongjae Lee and Seungwon Shin and Jin-Woo Chung},
   doi = {10.48550/arxiv.2204.06885},
   month = {4},
   title = {Shedding New Light on the Language of the Dark Web},
   url = {https://arxiv.org/abs/2204.06885v2},
   year = {2022},
}

@article{Blei2003LatentDirichletAllocation,
   author = {David M. Blei and Andrew Y. Ng and Michael I. Jordan},
   journal = {Journal of Machine Learning Research},
   pages = {993-1022},
   title = {Latent Dirichlet Allocation},
   volume = {3},
   year = {2003}
}

@misc{hugginface,
   author = {{Hugging Face}},
   title = {all-mpnet-base-v2},
   howpublished = {\url{https://huggingface.co/sentence-transformers/all-mpnet-base-v2}},
   month = {07},
   year = {2022}
}

@misc{all-mpnet-multilingual,
   author = {{Hugging Face}},
   title = {paraphrase-multilingual-mpnet-base-v2},
   howpublished = {\url{https://huggingface.co/sentence-transformers/paraphrase-multilingual-mpnet-base-v2}},
   month = {11},
   year = {2022}
}

@misc{ruBert,
   author = {{Hugging Face}},
   title = {rubert-base-cased},
   howpublished = {\url{https://huggingface.co/DeepPavlov/rubert-base-cased}},
   month = {11},
   year = {2022}
}

@misc{secBert,
   author = {{Hugging Face}},
   title = {SecBERT},
   howpublished = {\url{https://huggingface.co/jackaduma/SecBERT}},
   month = {11},
   year = {2022}
}

@inproceedings{reimers2020making,
  title={Making Monolingual Sentence Embeddings Multilingual using Knowledge Distillation},
  author={Reimers, Nils and Gurevych, Iryna},
  booktitle={Proceedings of the 2020 Conference on Empirical Methods in Natural Language Processing (EMNLP)},
  pages={4512--4525},
  year={2020}
}

@article{hili2014bios,
  title={The BIOS and Rootkits},
  author={Hili, Graham and Mayes, Keith and Markantonakis, Konstantinos},
  journal={Secure Smart Embedded Devices, Platforms and Applications},
  pages={369--381},
  year={2014},
  publisher={Springer}
}

@misc{googleTranslate,
title={What is Cloud Translation?},
author={Google},
howpublished={\url{https://cloud.google.com/translate/docs/overview}},
notes={[Online] Last accessed: October, 14 2022},
year={2023}
}

@misc{ncscglossary,
   author = {National Cyber Security Centre},
   title = {NCSC glossary - NCSC.GOV.UK},
   howpublished = {\url{https://www.ncsc.gov.uk/information/ncsc-glossary}},
   year = {2016},
}

@misc{frequencyOfEnglishPunctuation,
   author = {Vivian Cook},
   title = {Frequency of English Punctuation Marks},
   note={[Online] Last accessed. 24th May 2023},
   howpublished={\url{http://www.viviancook.uk/Punctuation/PunctFigs.htm}},
   year = {2023}
}

@article{averageWordLength,
  title={The average word length dynamics as an indicator of cultural changes in society},
  author={Bochkarev, Vladimir V and Shevlyakova, Anna V and Solovyev, Valery D},
  journal={Social Evolution and History},
  volume={14},
  number={2},
  pages={153--175},
  year={2015},
  publisher={Общество с ограниченной ответственностью" Издательство" Учитель"}
}

@inproceedings{Bhalerao2019mappingtheunderground,
  title={Mapping the underground: Supervised discovery of cybercrime supply chains},
  author={Bhalerao, Rasika and Aliapoulios, Maxwell and Shumailov, Ilia and Afroz, Sadia and McCoy, Damon},
  booktitle={2019 APWG Symposium on Electronic Crime Research (eCrime)},
  pages={1--16},
  year={2019},
  organization={IEEE}
}

@inproceedings{Nunes2016darknetanddeepnet,
  title={Darknet and deepnet mining for proactive cybersecurity threat intelligence},
  author={Nunes, Eric and Diab, Ahmad and Gunn, Andrew and Marin, Ericsson and Mishra, Vineet and Paliath, Vivin and Robertson, John and Shakarian, Jana and Thart, Amanda and Shakarian, Paulo},
  booktitle={2016 IEEE Conference on Intelligence and Security Informatics (ISI)},
  pages={7--12},
  year={2016},
  organization={IEEE}
}

@article{rahman2021attackers,
  title={What are the attackers doing now? Automating cyberthreat intelligence extraction from text on pace with the changing threat landscape: A survey},
  author={Rahman, Md Rayhanur and Mahdavi-Hezaveh, Rezvan and Williams, Laurie},
  journal={ACM Computing Surveys},
  year={2021},
  publisher={ACM New York, NY}
}

@inproceedings{lusthaus2020mapping,
  title={Mapping the geography of cybercrime: A review of indices of digital offending by country},
  author={Lusthaus, Jonathan and Bruce, Miranda and Phair, Nigel},
  booktitle={2020 IEEE European Symposium on Security and Privacy Workshops (EuroS\&PW)},
  pages={448--453},
  year={2020},
  organization={IEEE}
}

@article{wagner2019cyber,
  title={Cyber threat intelligence sharing: Survey and research directions},
  author={Wagner, Thomas D and Mahbub, Khaled and Palomar, Esther and Abdallah, Ali E},
  journal={Computers \& Security},
  volume={87},
  pages={101589},
  year={2019},
  publisher={Elsevier}
}

@article{rand1971objective,
  title={Objective criteria for the evaluation of clustering methods},
  author={Rand, William M},
  journal={Journal of the American Statistical association},
  volume={66},
  number={336},
  pages={846--850},
  year={1971},
  publisher={Taylor \& Francis}
}

@misc{appendix,
title = {Appendix for: Investigating and Comparing Discussion Topics in Multilingual Underground Forums},
author = {anonymous for review},
year = {2025},
howpublished = {\url{https://github.com/mkmisch/pockets_of_knowledge}}
}

@INPROCEEDINGS{Bermudez2021shadyeconomy,
  author={Bermudez-Villalva, Adrian and Stringhini, Gianluca},
  booktitle={APWG Symposium on eCrime}, 
  title={The shady economy: Understanding the difference in trading activity from underground forums in different layers of the Web}, 
  year={2021},
  volume={},
  number={},
  pages={1-10},
  doi={10.1109/eCrime54498.2021.9738751}}

@article{bermudez2018under,
  title={Under and over the surface: a comparison of the use of leaked account credentials in the Dark and Surface Web},
  author={Bermudez Villalva, Dario Adriano and Onaolapo, Jeremiah and Stringhini, Gianluca and Musolesi, Mirco},
  journal={Crime Science},
  volume={7},
  number={1},
  pages={1--11},
  year={2018},
  publisher={Springer}
}

@inproceedings{edwards2018geography,
  title={The geography of online dating fraud},
  author={Edwards, Matthew and Suarez-Tangil, Guillermo and Peersman, Claudia and Stringhini, Gianluca and Rashid, Awais and Whitty, Monica},
  booktitle={Workshop on technology and consumer protection},
  year={2018},
  organization={IEEE-TCSP}
}

@article{Iqbal2023Nextdoor, 
title={Lady and the Tramp Nextdoor: Online Manifestations of Real-World Inequalities in the Nextdoor Social Network}, 
volume={17}, 
url={https://ojs.aaai.org/index.php/ICWSM/article/view/22155}, 
DOI={10.1609/icwsm.v17i1.22155}, 
abstractNote={From health to education, income impacts a huge range of life choices. Earlier research has leveraged data from online social networks to study precisely this impact. In this paper, we ask the opposite question: do different levels of income result in different online behaviors? We demonstrate it does. We present the first large-scale study of Nextdoor, a popular location-based social network. We collect 2.6 Million posts from 64,283 neighborhoods in the United States and 3,325 neighborhoods in the United Kingdom, to examine whether online discourse reflects the income and income inequality of a neighborhood. We show that posts from neighborhoods with different incomes indeed differ, e.g. richer neighborhoods have a more positive sentiment and discuss crimes more, even though their actual crime rates are much lower. We then show that user-generated content can predict both income and inequality. We train multiple machine learning models and predict both income (R2=0.841) and inequality (R2=0.77).}, number={1}, 
journal={Proceedings of the International AAAI Conference on Web and Social Media}, author={Iqbal, Waleed and Ghafouri, Vahid and Tyson, Gareth and Suarez-Tangil, Guillermo and Castro, Ignacio}, 
year={2023}, 
month={Jun.}, 
pages={399-410} }

@INPROCEEDINGS{Zhao2016jargon,
  author={Zhao, Kangzhi and Zhang, Yong and Xing, Chunxiao and Li, Weifeng and Chen, Hsinchun},
  booktitle={2016 IEEE Conference on Intelligence and Security Informatics (ISI)}, 
  title={Chinese underground market jargon analysis based on unsupervised learning}, 
  year={2016},
  volume={},
  number={},
  pages={97-102},
  doi={10.1109/ISI.2016.7745450}}

@article{Li2022unstructured,
title = {Neural Natural Language Processing for unstructured data in electronic health records: A review},
journal = {Computer Science Review},
volume = {46},
pages = {100511},
year = {2022},
issn = {1574-0137},
doi = {https://doi.org/10.1016/j.cosrev.2022.100511},
url = {https://www.sciencedirect.com/science/article/pii/S1574013722000454},
author = {Irene Li and Jessica Pan and Jeremy Goldwasser and Neha Verma and Wai Pan Wong and Muhammed Yavuz Nuzumlalı and Benjamin Rosand and Yixin Li and Matthew Zhang and David Chang and R. Andrew Taylor and Harlan M. Krumholz and Dragomir Radev}
}

@article{lochmiller2021conducting,
  title={Conducting thematic analysis with qualitative data},
  author={Lochmiller, Chad R},
  journal={The Qualitative Report},
  volume={26},
  number={6},
  pages={2029--2044},
  year={2021}
}

@article{birks2020unsupervised,
  title={Unsupervised identification of crime problems from police free-text data},
  author={Birks, Daniel and Coleman, Alex and Jackson, David},
  journal={Crime Science},
  volume={9},
  number={1},
  pages={18},
  year={2020},
  publisher={Springer}
}

@article{ruellan2024conti,
  title={Conti Inc.: understanding the internal discussions of a large ransomware-as-a-service operator with machine learning},
  author={Ruellan, Estelle and Paquet-Clouston, Masarah and Garcia, Sebasti{\'a}n},
  journal={Crime Science},
  volume={13},
  number={1},
  pages={16},
  year={2024},
  publisher={Springer}
}


\end{document}